\pdfoutput=1
\documentclass[a4paper]{PoS}

\usepackage{amsmath,amssymb}
\usepackage{subcaption}
\usepackage[export]{adjustbox}

\title{Dark Sectors from the  Hidden Photon Perspective}

\ShortTitle{Hidden Photon Dark Sectors}

\author{\speaker{Patrick Foldenauer} \\ 
        Institut f\"ur Theoretische Physik, Universit\"at Heidelberg, Philosophenweg 16, 69120 Heidelberg, Germany\\
        E-mail: \email{foldenauer@thphys.uni-heidelberg.de}}

\abstract{The non-observation of dark matter (DM) by direct detection experiments suggests that any new interaction of DM with the Standard Model (SM) should be very weak.
 One of the simplest scenarios to achieve this is a dark sector that is charged under a new  $U(1)_X$ symmetry, which is kinetically mixed with the SM hypercharge $U(1)_Y$. We briefly review  the status of such a minimal setup and analyze in a second step how the picture is altered if also SM fields are charged under the new symmetry. We exemplify this for the case of a gauged  $U(1)_{L_\mu-L_\tau}$ and show that this allows for a simultaneous explanation of the $(g-2)_\mu$ excess and the DM relic abundance $\Omega_{DM}$.  Furthermore, we discuss the potential of four-lepton and two-lepton plus missing energy signatures to test such scenarios.}

\FullConference{7th Annual Conference on Large Hadron Collider Physics - LHCP2019\\
		20-25 May, 2019\\
		Puebla, Mexico}

\begin{document}

\section{Introduction}
{\textit{The majority of the material presented here is based on \cite{Bauer:2018onh,Foldenauer:2018zrz} and partly on \cite{FoldenauerPhd}, which we refer the reader to for detailed discussions and an expanded list of references.}}\par
Theoretical particle physics today can be viewed as being in a peculiar position.
On the one hand, the formulation of the SM in terms of a renormalizable quantum field theory invariant under local symmetry transformations was a major leap towards a unified description of the fundamental interactions of particles.  
Only by virtue of this framework it has become possible to
predict the anomalous magnetic moment $(g-2)_e$ of the electron with an accuracy of better than $10^{-12}$ \cite{Hanneke:2008tm}. 
On the other hand, however, we know from cosmological and astrophysical observations that the bulk of the energy of the universe is abundant in the form of dark matter (DM) and dark energy~\cite{Aghanim:2018eyx}, which are not described by SM physics. Furthermore, the SM alone can most likely not provide enough CP-violation to produce the baryon asymmetry leading to the observed baryon relic abundance accounting for roughly $5\%$ of the energy budget of the universe. In particular, the fact that there is roughly five times as much dark matter as baryonic matter motivates for the presence of a dark sector that interacts with the SM only very weakly. \par
While this serves as a clear motivation to consider extensions of the SM, we do not want to give up on the very successful concepts of renormalizability and gauge invariance.
 A systematic analysis of bottom-up extensions of the SM via operators coupling SM degrees of freedom to new dark sector fields shows that there are only three types of gauge invariant, (perturbatively) renormalizable operators. These are the so-called Higgs, neutrino and vector portal interactions, which all mediate interactions of the SM with a potential new dark sector. Of these three the vector portal interaction requires the presence of a new Abelian gauge group, $U(1)_X$, thereby extending the symmetry structure of the SM. As the field strength tensor $X_{\mu\nu}$ of an Abelian $U(1)_X$ symmetry is gauge invariant by itself, this allows for the kinetic mixing term  \cite{Okun:1982xi,Holdom:1985ag},
 \begin{equation}
\mathcal{L} = - \frac{\epsilon_Y}{2} B_{\mu\nu}X^{\mu\nu} \,, \label{eq:vecportal}
\end{equation}
 coupling the field strength $X_{\mu\nu}$ of the new symmetry to that of SM hypercharge,  $B_{\mu\nu}$. \par
 
Such extra secluded $U(1)_X$ symmetries (under which SM fields remain uncharged) rejoiced from an elevated popularity in the context  of (supersymmetric) DM models \cite{ArkaniHamed:2008qn,ArkaniHamed:2008qp,Hooper:2008im,Baumgart:2009tn,Cheung:2009qd,Katz:2009qq,Morrissey:2009ur} when it was realized that they could naturally help to explain an excess in the observed cosmic-ray positron fraction. Accordingly, in the literature this has lead to an increased interest in searches for such extra $U(1)_X$ symmetries~\cite{Batell:2009yf,Essig:2009nc,Reece:2009un,Bjorken:2009mm}.\par
 
 In the following, we want to use this as motivation to briefly review the current status of such a minimal secluded $U(1)_X$ symmetries. Going beyond the minimal case, we will investigate how the phenomenology of such models is changed if also SM fields are charged under the new symmetry. In particular, we show that in a gauged $U(1)_{L_\mu-L_\tau}$ an explanation of the $(g-2)_\mu$ excess~\cite{Bennett:2006fi} is still viable. Furthermore, we show that  under addition of a vector-like fermion $\chi$ this can be supplemented by a simultaneous explanation of the observed DM relic abundance $\Omega_{DM}$. Finally,  we will discuss experimental strategies to test such scenarios in the future.

\section{Hidden photons in a nutshell}

A minimal choice of extending the SM gauge group $G_\mathrm{SM}$ is to add a new secluded $U(1)_X$ under which SM fields remain uncharged. The minimal setup is given by the Lagrangian 
\begin{equation}\label{eq:min_lagr}
\mathcal{L} = \mathcal{L}_\mathrm{SM} -\frac{1}{4} X_{\mu\nu} X^{\mu\nu} - \frac{\epsilon_Y}{2} B_{\mu\nu} X^{\mu\nu} - \frac{M_X^2}{2} X_{\mu}X^{\mu} - g_x\, j_x^\mu X_{\mu} \,,
\end{equation}
where $X_{\mu}$ and  $X_{\mu\nu}$ denote the new gauge boson and associated field strength tensor. In principle, there can be a conserved current of particles carrying charge under the new $U(1)_X$ symmetry, 
\begin{equation} \label{eq:protocurrent}
j^\mu_x = \sum_\psi \bar \psi\, \gamma^\mu\,(Q^\psi_L\, P_L + Q^\psi_R\, P_R)\, \psi\,.
\end{equation}
\begin{figure}[t]
\begin{center}
\includegraphics[width=.7\textwidth]{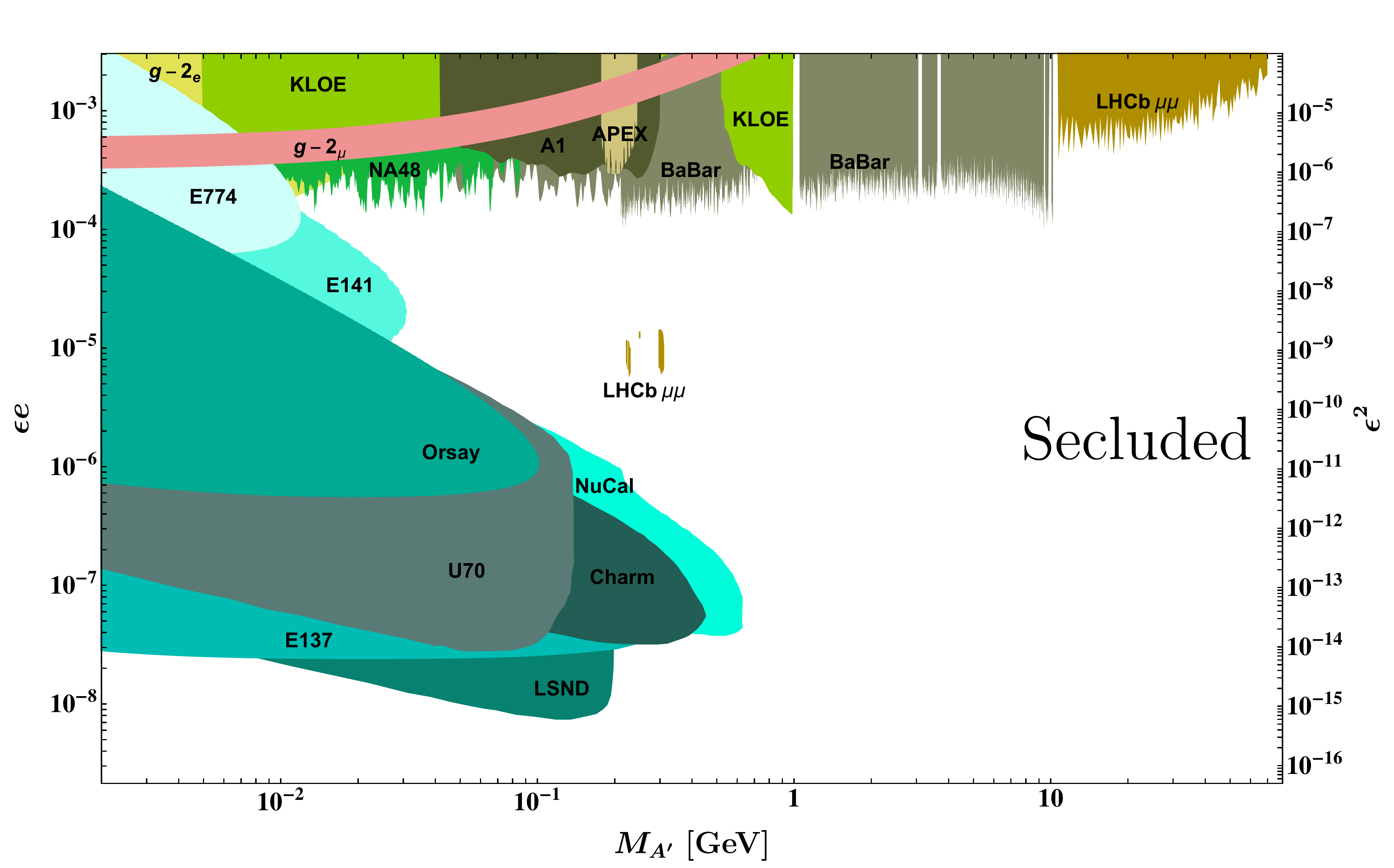}
\includegraphics[width=.7\textwidth]{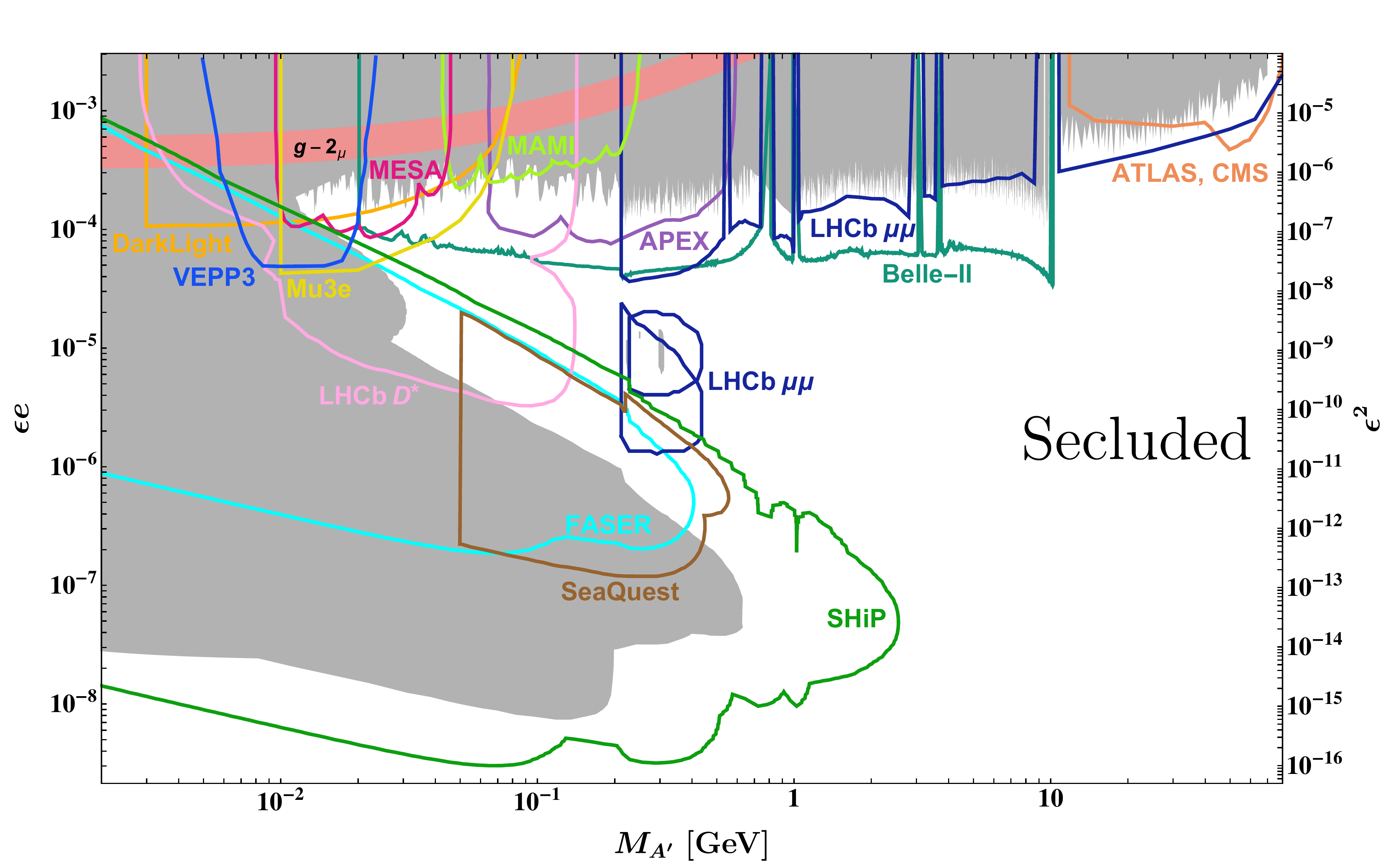}
\end{center}
\caption{\label{fig:lims_sec}  Limits on a $U(1)_X$ gauge boson with mass $M_{A'}$ and kinetic mixing $\epsilon$. The red bands show the $2\sigma$ preferred region for $(g-2)_\mu$. (Top) Existing constraints. (Bottom) Projections of future planned experiments. Figures taken from~\cite{Bauer:2018onh}.}
\end{figure}
%
However, in the minimal setup (the secluded case) we assume that there are no field $\psi$ charged under the new symmetry\footnote{In a more realistic setup there will be some new heavy fields $\psi$ potentially both charged under the SM and $U(1)_X$. At the electroweak scale these will, however, have completely decoupled and can be integrated out. This will in turn induce a kinetic mixing  term between the hypercharge boson $B_\mu$ and the new $U(1)_X$ boson $X_\mu$ from loop effects.} and we take the current to be zero, $j_\mu^x=0$. The gauge boson mass $M_X$ can be generated from a Higgs mechanism if there is an additional new scalar $\phi$ charged under $U(1)_X$ that attains a VEV and breaks the new symmetry. However, the most economical way is to generate $M_X$ via a  St\"uckelberg mechanism. There a new pseudo-scalar field $\sigma$ obeys simultaneous transformation rules with the vector boson $X_\mu$ under a $U(1)_X$ symmetry transformation and makes up the longitudinal component of $X_\mu$. In this setup the gauge boson mass term is gauge-invariant and $M_X$ is  a free parameter of the theory. \par
Due to the presence of the kinetic mixing term, the Lagrangian \eqref{eq:min_lagr} is not properly canonically normalized. After a field redefinition of the neutral gauge bosons to properly normalize the kinetic terms  and a rotation to the  mass basis,  we obtain the interactions of the fermions with the physical mass eigenstates of the gauge bosons as
\begin{align}\label{eq:currentcouplings}
{\mathcal{L}}_\mathrm{int}=&\left(ej_\text{EM}^\mu , \frac{e}{\sin \theta_w \cos \theta_w} j_Z^\mu, g_{x}j_{x}^\mu\right) \,K\,\begin{pmatrix}A_\mu\\ Z_\mu\\  A'_\mu\end{pmatrix} \,, 
\end{align}
where $A'_\mu$ is the corresponding mass eigenstate of the $U(1)_X$ boson and to leading order
\begin{align}
K= \begin{pmatrix}
1 & 0 & -\epsilon \phantom{e}\\
0 & 1& 0 \\
0 & \epsilon \tan\theta_w&  1
\end{pmatrix} \,,
\label{eq:KK}
\end{align}
with the definition $\epsilon=\epsilon_Y\cos(\theta_w)$.
To first approximation, also the mass of the new boson before and after the basis change are equal,
\begin{equation}
M_{A'}^2={M}_{X}^2(1 +\mathcal{O}(\epsilon^2)).
\end{equation}

From \eqref{eq:currentcouplings} it can be seen that in absence of a new gauge current ($j_x^\mu=0$) the only interaction of the $A'$ with the SM is via the kinetic mixing suppressed coupling  to  the electromagnetic current $j_\text{EM}^\mu$. This naturally suggests the name \textbf{hidden} or \textbf{dark photon} for the mass eigenstate $A'$. Due to the suppression of the interaction with $\epsilon$, the production of an $A'$ in electromagnetic interactions is a rare process. Therefore, these particles are typically searched for in experiments with a large number of electromagnetic interactions and good control of the backgrounds.
Two general classes of experiments where these conditions are met are collider experiments on the one side and beam dump and fixed target experiments on the other side. As can be seen in Figure~\ref{fig:lims_sec} collider experiments are sensitive to moderate kinetic mixings $\epsilon$ in a very broad mass range (top region in the plots). On the other hand, beam dump and fixed target experiments are sensitive to much smaller couplings in the small mass region (left region in the plots).

\section{Beyond the minimal case}

Instead of  {\it ad hoc} postulating a new fundamental $U(1)_X$ symmetry, we can try to gauge one of the accidental global symmetries of the SM Lagrangian, $U(1)_{B}$, $U(1)_{L_e}$, $U(1)_{L_\mu}$ or $U(1)_{L_\tau}$. In fact, the four combinations $U(1)_{L_\mu-L_e}$, $U(1)_{L_e-L_\tau}$, $U(1)_{L_\mu-L_\tau}$ and $U(1)_{B-L}$ (under addition of three right-handed neutrinos) are anomaly-free with only the SM field content. 

\begin{figure}[t]
\begin{center}
\includegraphics[width=.7\textwidth]{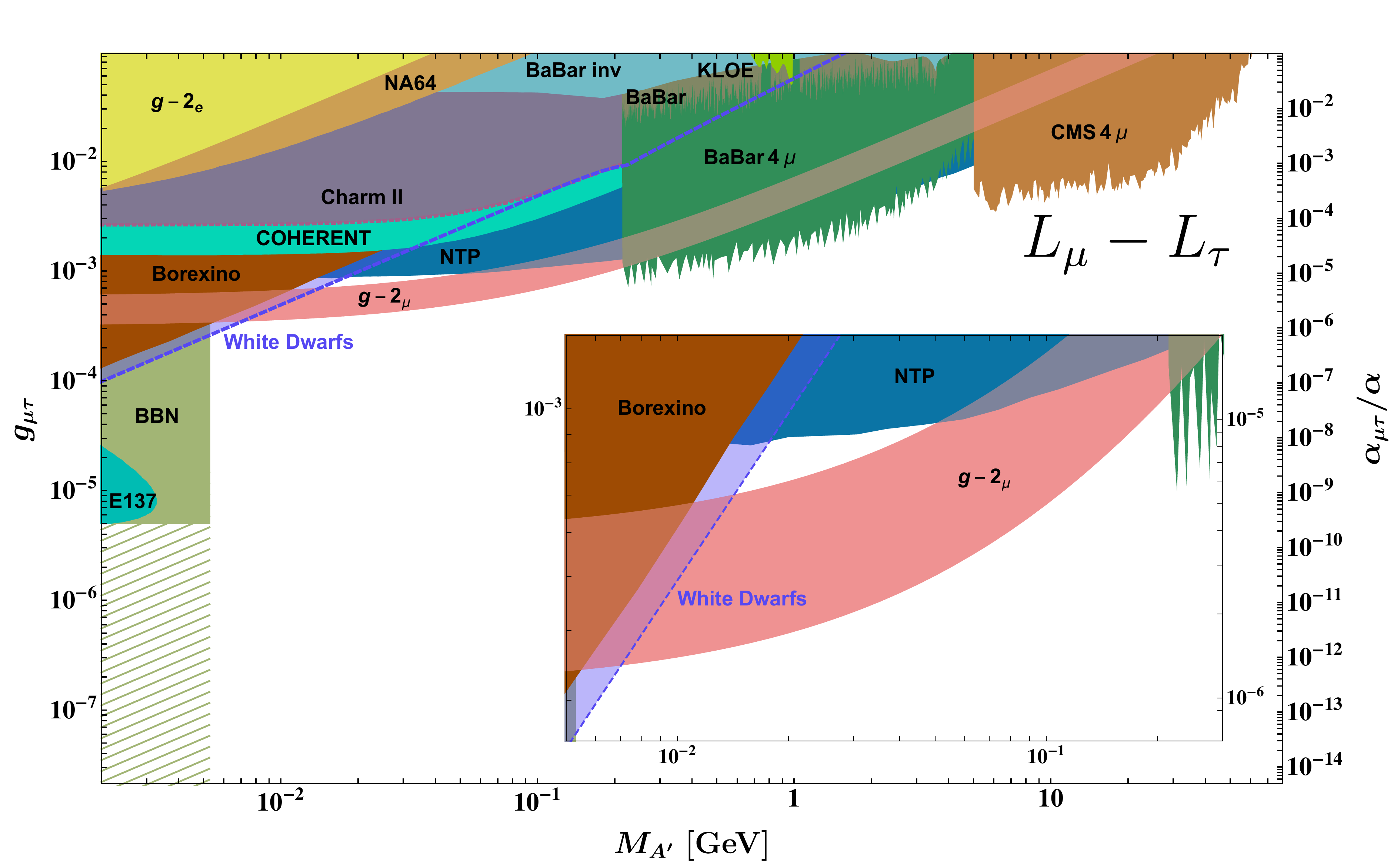}
\includegraphics[width=.7\textwidth]{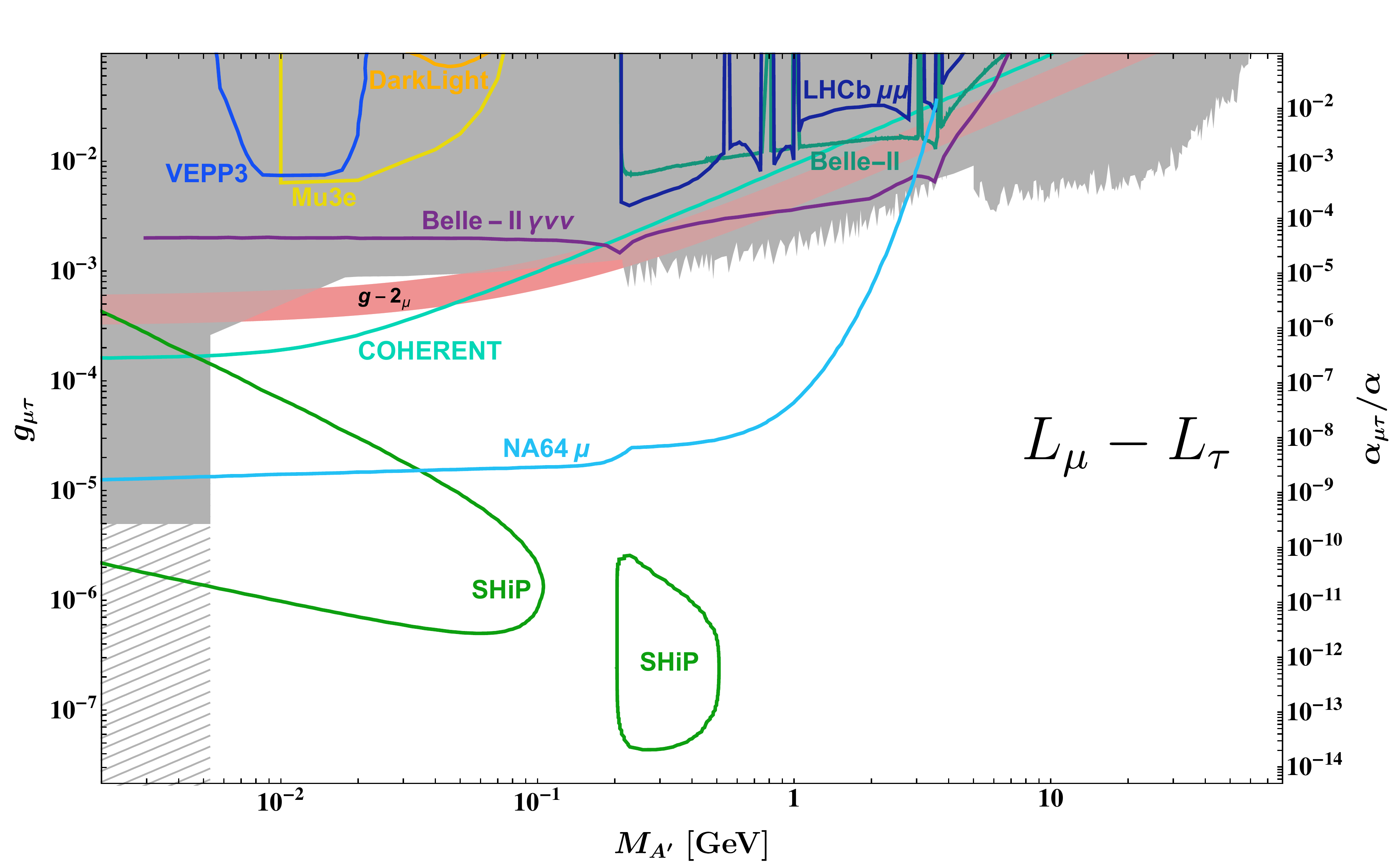}
\end{center}
\caption{\label{fig:lims_mutau}  Limits on a $U(1)_{L_\mu-L_\tau}$ gauge boson with mass $M_{A'}$ and coupling constant $g_{\mu\tau}$. The red bands show the $2\sigma$ preferred region for $(g-2)_\mu$. (Top) Existing constraints. (Bottom) Projections of future planned experiments. Figures adapted from~\cite{Bauer:2018onh} (CMS 4$\mu$ bound added).}
\end{figure}
%

In  these scenarios the new gauge boson  will couple to  the  gauge currents of SM fermions 
\begin{align}
j_{\, i-j}^\mu&= \bar L_i \gamma^\mu L_i 
          + \bar \ell_{R,i}\gamma^\mu \ell_{R,i} 
          - \bar L_j \gamma^\mu L_j -\bar\ell_{R,j}\gamma^\mu \ell_{R,j}\,,
            \qquad && U(1)_{L_i-L_j} \,,\notag \\ 
j_{B-L}^\mu&=  \frac{1}{3}\bar Q \gamma^\mu Q 
          + \frac{1}{3}\bar u_R\gamma^\mu u_R 
          + \frac{1}{3}\bar d_R\gamma^\mu d_R
          - \bar L \gamma^\mu L 
          - \bar \ell_R\gamma^\mu \ell_R
          - \bar \nu_R\gamma^\mu \nu_R\,,
            \qquad && U(1)_{B-L} \;.
\label{eq:fermion_currents}
\end{align}
As there are now particles charged both under  hypercharge  and the new $U(1)$ symmetry, kinetic mixing will be induced at the loop level. In the case of the groups $U(1)_{L_i-L_j}$ the loop-induced mixing is finite and can be calculated from 
\begin{align}\label{eq:lmutaumix}
\epsilon_{ij}(q^2)=-\frac{e\, g_{ij}}{4\pi^2}\int_0^1 dx\,x(x-1)\bigg[3\log\bigg(\frac{m_{\ell_i}^2+q^2x(x-1)}{m_{\ell_j}^2+q^2x(x-1)}\bigg)+ \log\bigg(\frac{m_{\nu_i}^2+q^2x(x-1)}{m_{\nu_j}^2+q^2x(x-1)}\bigg) \bigg]\,.
\end{align}
As  a consequence, these models expose  a strong hierarchy in the couplings of the hidden photon $A'$ to SM fermions. Interactions due to loop-induced mixing are suppressed compared to tree-level gauge interactions by a factor of 
\begin{equation}
\frac{\epsilon_{ij}(q^2)\, e}{g_{ij}} \propto \frac{\alpha}{\pi} \log\left(\frac{m_{\ell_j}}{m_{\ell_i}}\right)\approx\mathcal{O}(10^{-2})\,.
\end{equation}
In \cite{Bauer:2018onh} we have reanalyzed the limits on secluded hidden photons shown in Figure~\ref{fig:lims_sec} in the context of these gauge groups and compared them to additional constraints not taken into account in the minimal secluded case, as e.g. from neutrino interactions.

\subsection{Example: $U(1)_{L_\mu-L_\tau}$}

 The induced coupling hierarchy has particularly dire consequences for the hidden photon limits in the case of $U(1)_{L_\mu-L_\tau}$,  as illustrated in Figure~\ref{fig:lims_mutau}. The corresponding gauge boson only has gauge interaction with second and third generation leptons. However, most  experimental setups rely on proton or electron beams for particle production. Hence, the production of  $U(1)_{L_\mu-L_\tau}$ bosons proceeds via loop-suppressed kinetic mixing interactions. This is the reason why most hidden photon limits from collider and beam dump experiments are a lot less stringent or absent entirely in this case.
On the other hand side, under all four gauge groups, $U(1)_{L_i-L_j}$ and $U(1)_{B-L}$,  neutrinos are now charged leading to unsuppressed interactions with the corresponding gauge bosons. This makes elastic neutrino scattering (Borexino, COHERENT, Charm-II), neutrino trident production (NTP) and non-standard interactions (NSI) sensitive probes for this class of hidden photons. These constraints are particularly important for $U(1)_{L_\mu-L_\tau}$ bosons with masses below the dimuon threshold. In this region the $A'$ can to good approximation only decay into neutrinos and hence not be tested in visible final state searches. This situation is summarized in the upper panel of Figure~\ref{fig:lims_mutau}.
\par
Most interestingly,  the observed excess of the anomalous magnetic moment of the muon \mbox{$(g-2)_\mu$} can still be explained in a gauged $U(1)_{L_\mu-L_\tau}$ for boson masses of 7 MeV $\lesssim M_{A'} \lesssim 200$ MeV. This is emphasized by the inset plot in the upper panel  in Figure~\ref{fig:lims_mutau}, where the red band shows the preferred $2\sigma$ region of the observed $(g-2)_\mu$ excess. As shown in the lower panel of Figure~\ref{fig:lims_mutau},  this region can be tested  in the future with the full COHERENT data set and potentially with a dedicated muon run of NA64.

\subsection{Four lepton searches}

\begin{figure}[t]
\centering
\begin{subfigure}{.4\textwidth}
  \centering
 \adjincludegraphics[ trim={0 0 0 0},clip, width=\textwidth]{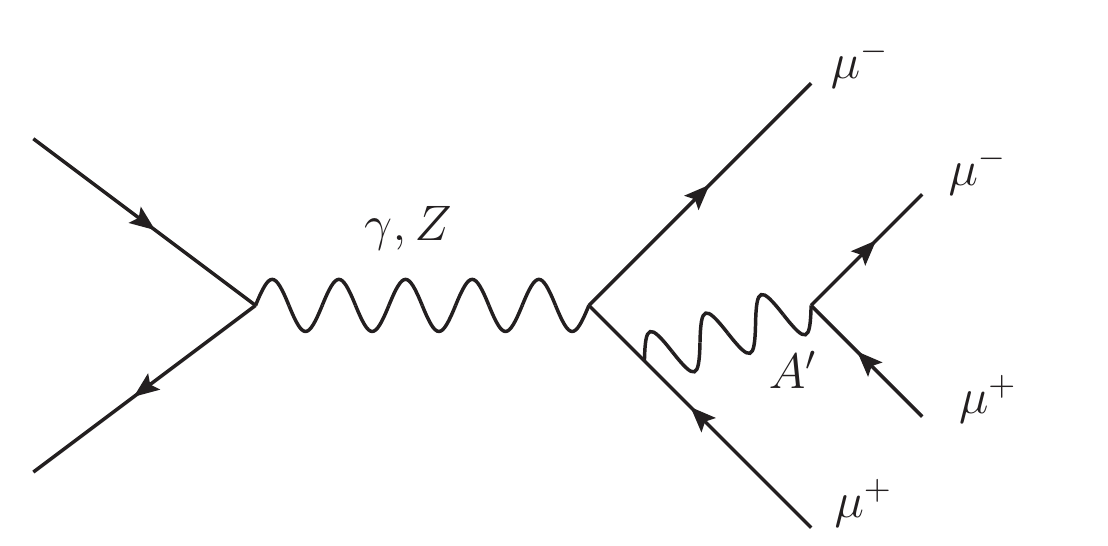}
\end{subfigure} %
\begin{subfigure}{.4\textwidth}
  \centering
   \adjincludegraphics[ trim={0 0 0 0},clip, width=\textwidth]{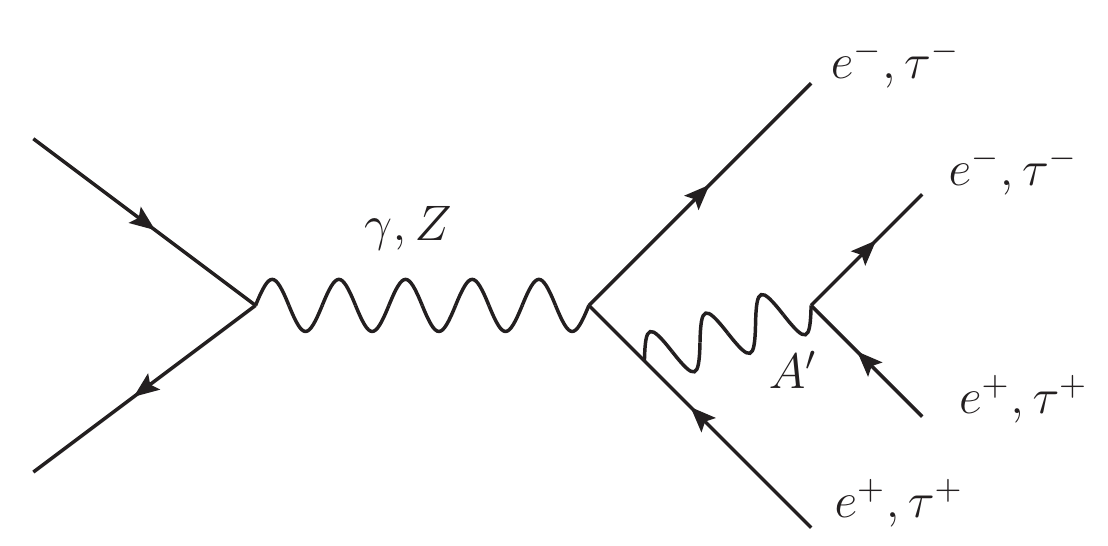}
\end{subfigure} 
 \caption{Tree-level diagrams of lepton-coupled hidden photons contributing to four-lepton final states. (Left) Four-muon final state e.g. in gauged $U(1)_{L_\mu-L_\tau}$. (Right) Potential $e^+e^-\tau^+\tau^-$, four-electron and four-tau signature in a gauged $U(1)_{L_e-L_\tau}$.}
\label{fig:4lep}
\end{figure}

Particularly  promising signatures to test models of gauged lepton-family number differences $U(1)_{L_i-L_j}$  are four-lepton final states.
For example, above the dimuon threshold $M_{A'} > 2 m_\mu$ the  $U(1)_{L_\mu-L_\tau}$ boson is most constrained  from searches for four-muon final states due to the process illustrated in the left panel of Figure~\ref{fig:4lep}. Searches for such signatures have been conducted by BaBar and CMS\footnote{Note that the CMS four-muon constraint~\cite{Sirunyan:2018nnz}  has appeared after the original publication of~\cite{Bauer:2018onh} and has been added here to the $U(1)_{L_\mu-L_\tau}$ limit plot for completeness.} and are shown in green and orange in  the upper panel of Figure~\ref{fig:lims_mutau}.  \par
The high sensitivity of  this signature is due to the fact that  the $U(1)_{L_\mu-L_\tau}$ boson can have a large contribution to four-muon final states as it is produced via an unsuppressed gauge interaction in final state radiation off a muon leg. Contrarily, the gauge boson has to to be produced via loop-suppressed kinetic mixing in Drell-Yan-like two-muon signatures making these significantly less sensitive.
\begin{figure}[t]
\begin{subfigure}{.5\textwidth}
  \centering
 \adjincludegraphics[ trim={0 0 0 0},clip, width=\textwidth]{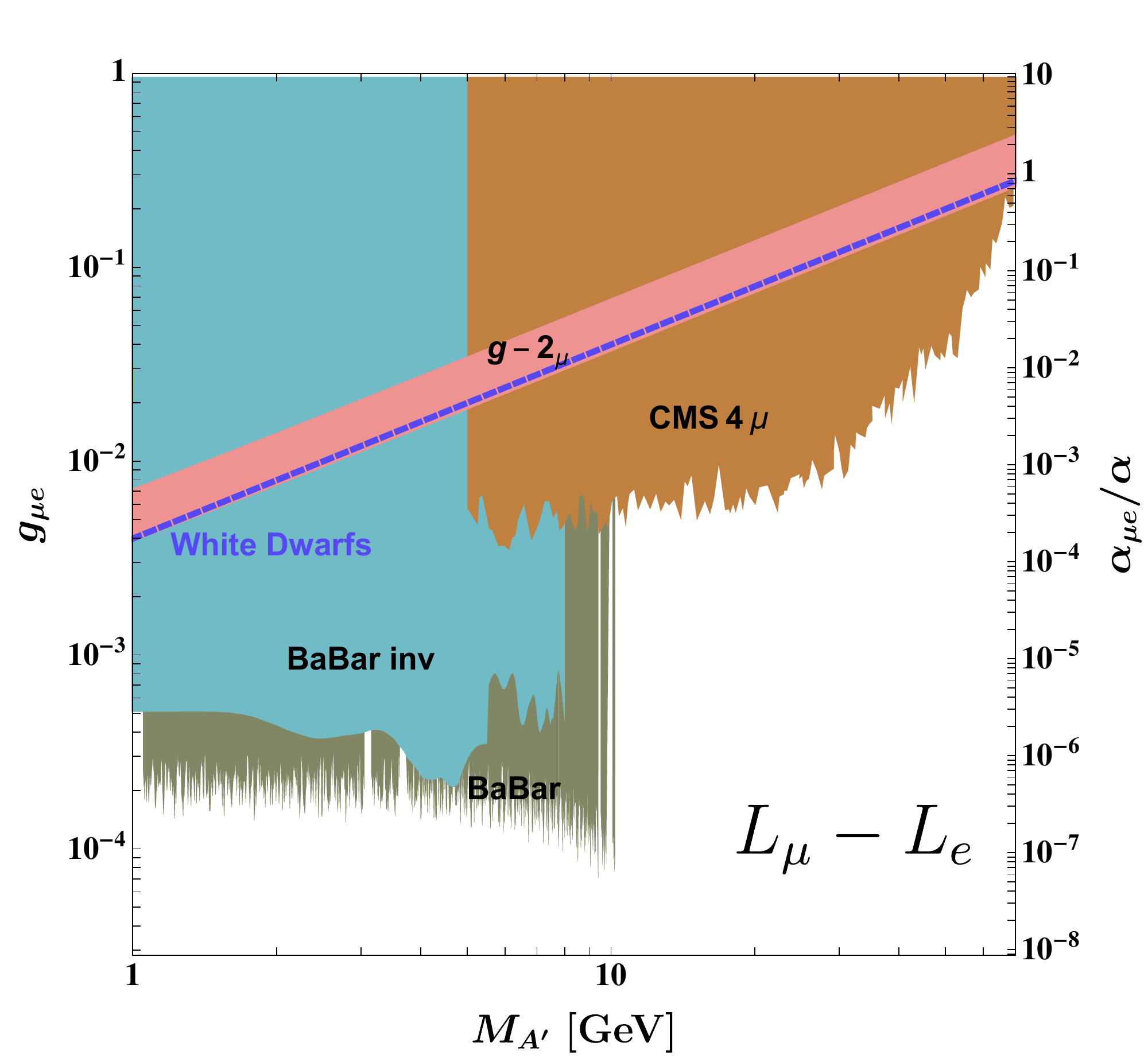}
\end{subfigure} %
\begin{subfigure}{.5\textwidth}
  \centering
   \adjincludegraphics[ trim={0 0 0 0},clip, width=\textwidth]{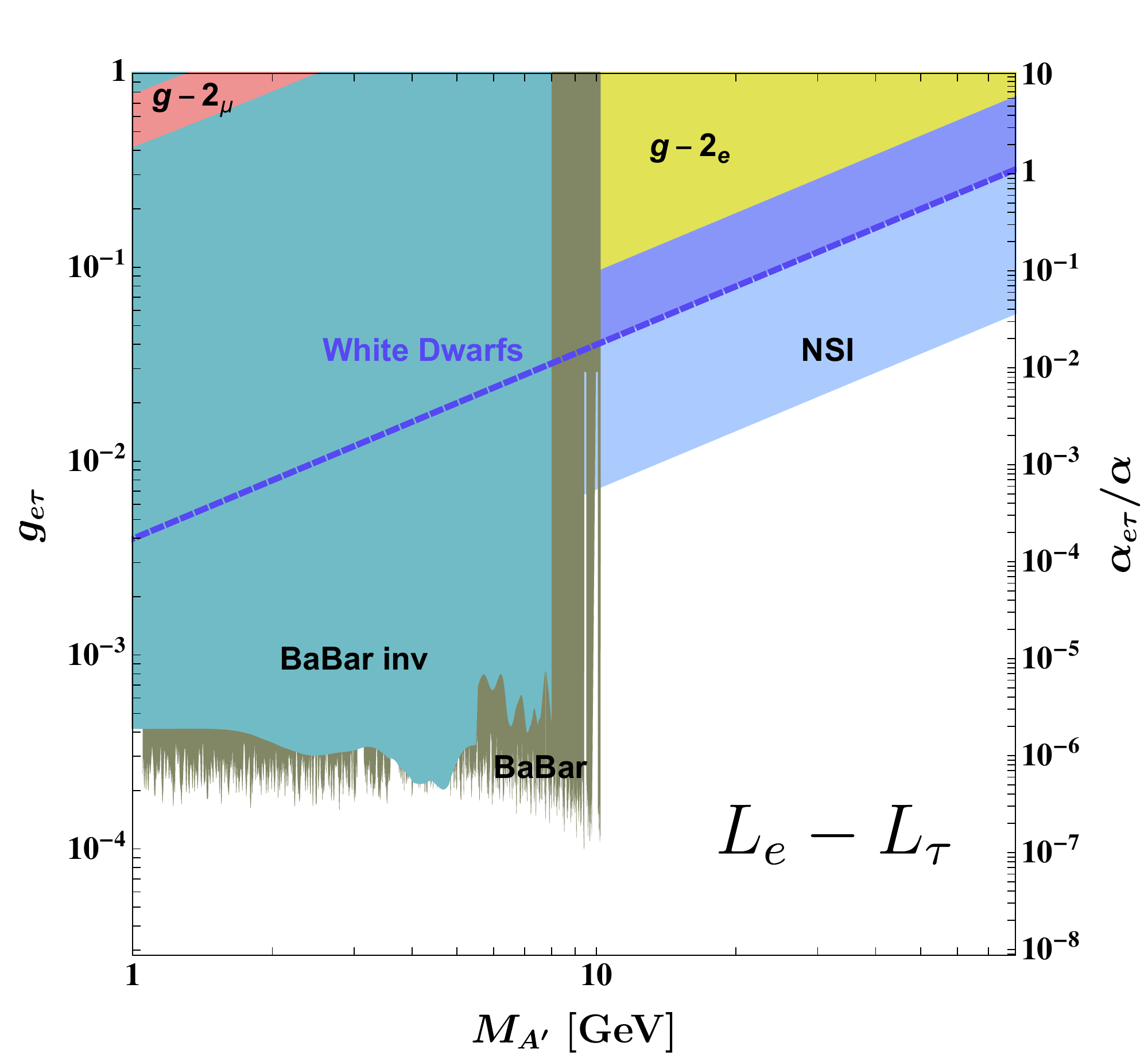}
\end{subfigure} 
 \caption{Limits on leptophilic hidden photons with GeV-scale masses. (Left) Limits on $U(1)_{L_\mu-L_e}$ bosons. The orange region shows the recasted constraint from four-muon searches of \cite{Sirunyan:2018nnz}. (Right) Limits on $U(1)_{L_e-L_\tau}$ bosons. The $(g-2)_e$ constraint is due to a recently improved determination of $\alpha$ \cite{Parker:2018vye} and is at the $3\sigma$ level. For high $A'$ masses the leading constraint comes from non-standard neutrino interactions (NSI) \cite{Heeck:2018nzc}.}
\label{fig:high_mass}
\end{figure}
In the case of $U(1)_{L_\mu-L_e}$ and $U(1)_{L_e-L_\tau}$, there are typically quite stringent limits on the associated hidden photons from BaBar dielectron searches for masses below the  $\Upsilon (4S)$ resonance  $M_{A'}\lesssim 10.5$ GeV (cf. Figure~\ref{fig:high_mass}). However, for hidden photons with higher masses these limits are absent as they cannot be produced on resonance anymore and the corresponding parameter space is much less constrained. \par
In analogy to the $U(1)_{L_\mu-L_\tau}$ case, we have recasted the limits from the  CMS four-muon search of \cite{Sirunyan:2018nnz} to the case of $U(1)_{L_\mu-L_e}$. The obtained constraint is shown in orange in the left panel of Figure~\ref{fig:high_mass} and interestingly excludes a high-mass explanation of $(g-2)_\mu$. In the case of a gauged $U(1)_{L_e-L_\tau}$ the four-muon constraint is absent as the corresponding hidden photon only couples to muons via loop-suppressed kinetic mixing. As illustrated in the right panel of Figure~\ref{fig:high_mass}, in this case the leading constraint is due to non-standard neutrino interactions (NSI). This bound has been derived in \cite{Heeck:2018nzc} and excludes hidden photons with  $M_{A'}/g_{e\tau} > 1.4$ TeV. 
Analogously to the CMS four-muon search, $U(1)_{L_e-L_\tau}$ bosons could be searched for e.g.~at ATLAS and CMS in $e^+e^-\tau^+\tau^-$, four-electron and four-tau final states due to the diagrams shown in the right panel of Figure~\ref{fig:4lep}. It would furthermore be interesting to evaluate the sensitivity of LHCb to the same processes if the $A'$ is long-lived and one of the lepton-pairs would originate from a displaced vertex. Such searches might in principle be sensitive to small gauge couplings in the high-mass region and could potentially improve over the NSI bound.

\section{Adding dark matter}

\begin{figure}[t]
\begin{center}
\includegraphics[width=.7\textwidth]{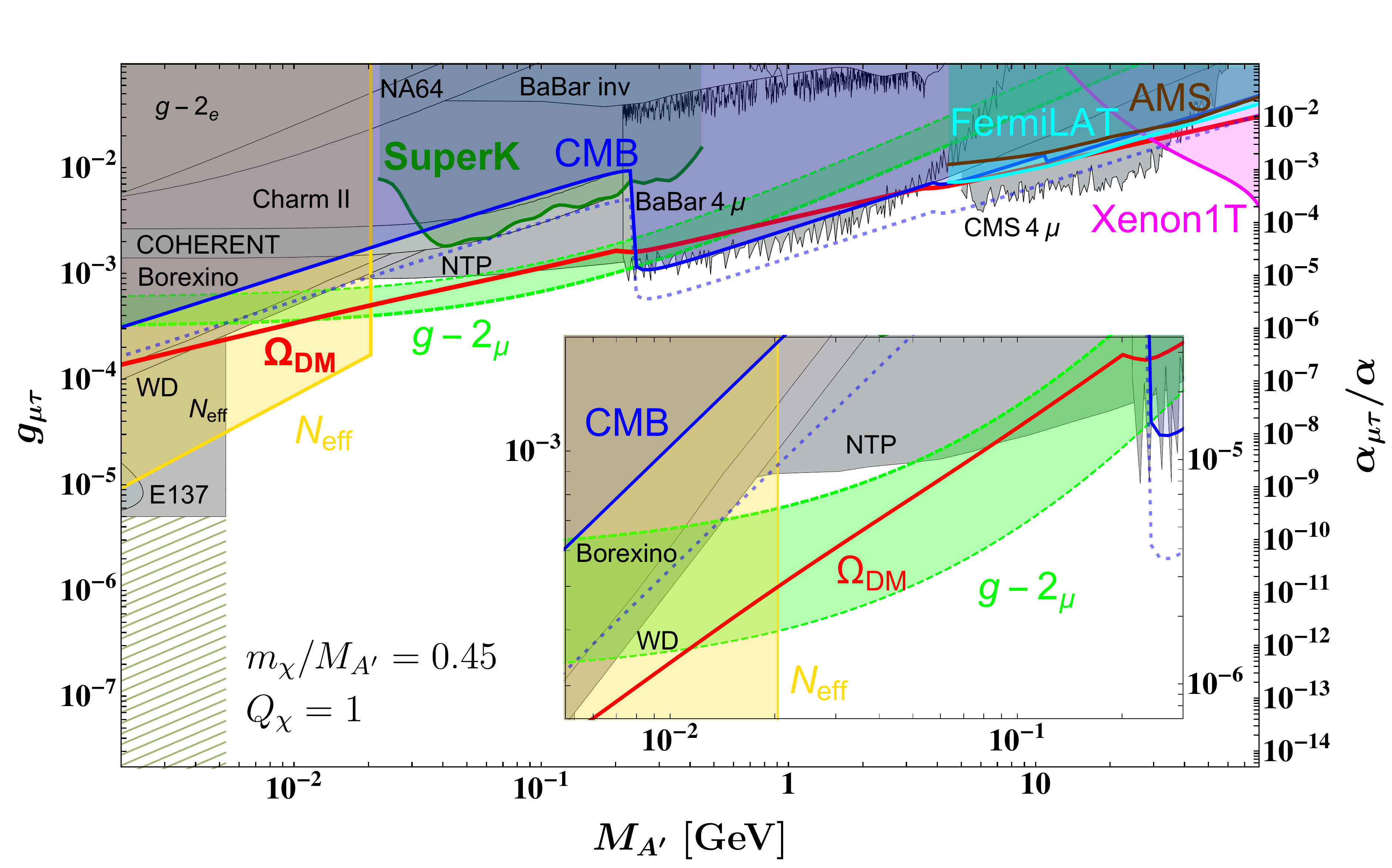}
\includegraphics[width=.7\textwidth]{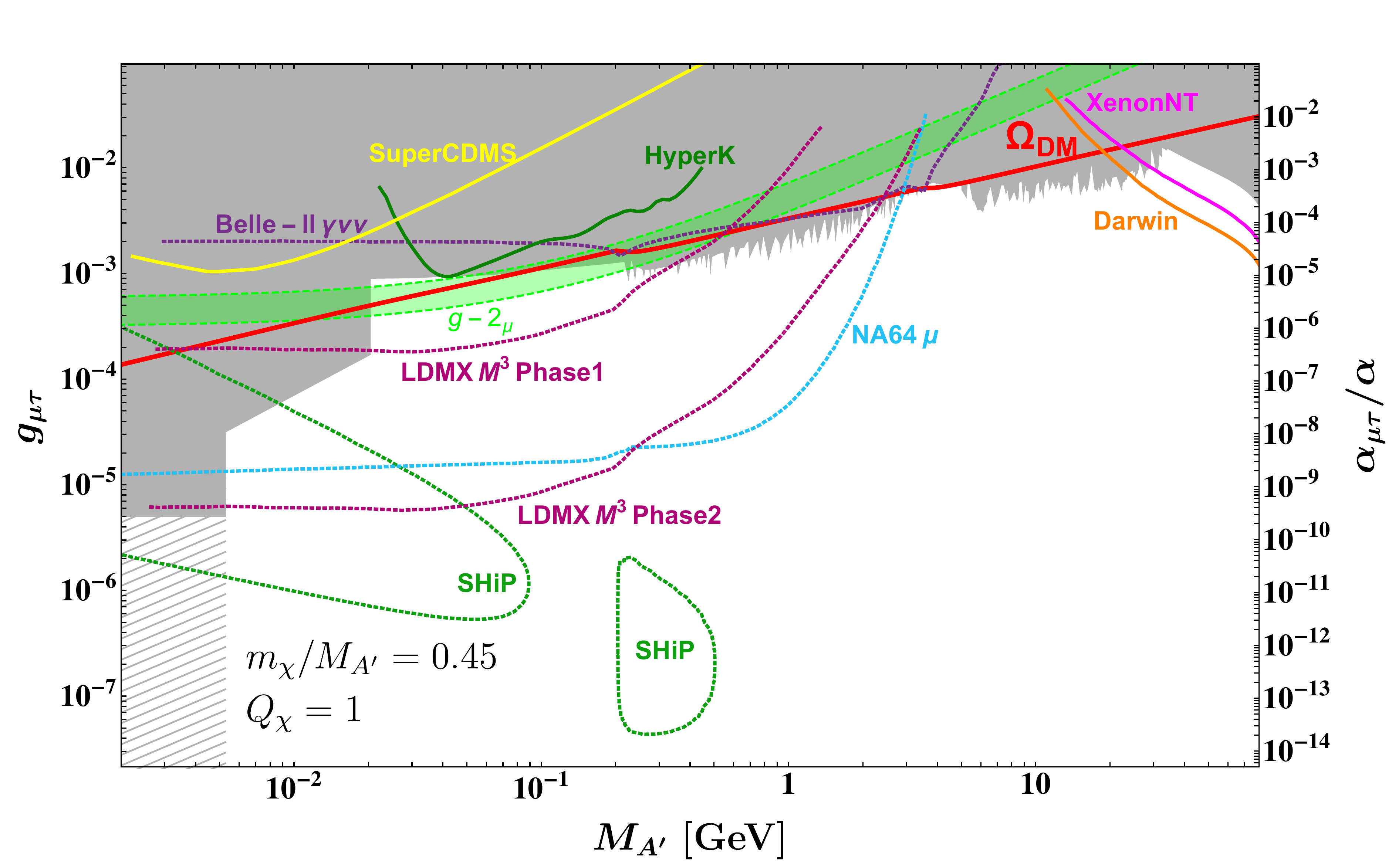}
\end{center}
\caption{\label{fig:lims_dm}  Limits on combined parameter space of the vector-like fermion $\chi$ and the $U(1)_{L_\mu-L_\tau}$ gauge boson in the coupling verusus mass plane. The green bands show the $2\sigma$ preferred region for $(g-2)_\mu$. The red curve shows the points where $\chi$ can explain all of the observed DM relic abundance $\Omega_\mathrm{DM}$. (Top) Existing constraints. (Bottom) Projections of future planned experiments. Figures taken from~\cite{Foldenauer:2018zrz}.}
\end{figure}
%

Thus far we have only discussed  the phenomenological consequences of extending the SM gauge group $G_\mathrm{SM}$. 
In a further step, we can consistently incorporate DM  
by adding a vector-like fermion $\chi$  carrying charge $Q_\chi$ under the new $U(1)$ symmetry,
\begin{align}
\mathcal{L}_\chi=-g_{x}\,  Q_\chi \, \bar \chi \gamma_\mu \chi \  X^\mu - m_\chi \bar \chi \chi  \,.
\end{align}
  This minimal choice preserves anomaly freedom of the gauge group. \par
For concreteness, we will once more consider the example of  $U(1)_{L_\mu-L_\tau}$. In Figure~\ref{fig:lims_dm} we show the constraints on the combined hidden photon and DM parameter space for a scenario with a fixed mass ratio $m_\chi/M_{A'}=0.45$  and DM charge $Q_\chi=1$. In the upper panel, the grey contours in the background are the current hidden photon constraints analogous to those shown in Figure~\ref{fig:lims_mutau}. The colored contours superimposed in the foreground are the current DM constraints on the vector-like fermion $\chi$. In the-low mass region the most important DM constraints are due to the number of effective neutrino degrees of freedom $N_\mathrm{eff}$ during big bang nucleosynthesis (BBN) and due to extra ionization during the cosmic microwave background (CMB). The red curve denotes the points where $\chi$ accounts for all of the observed DM relic abundance $\Omega_\mathrm{DM}$. As can be seen from the inset plot, in this scenario the observed $(g-2)_\mu$ excess and the DM abundance $\Omega_\mathrm{DM}$ can be simultaneously explained in the range of 20 MeV $\lesssim M_{A'} \lesssim 100$ MeV.\par
\begin{figure}[t]
\begin{subfigure}{.45\textwidth}
 \adjincludegraphics[ trim={0 0 0 0},clip, width=\textwidth]{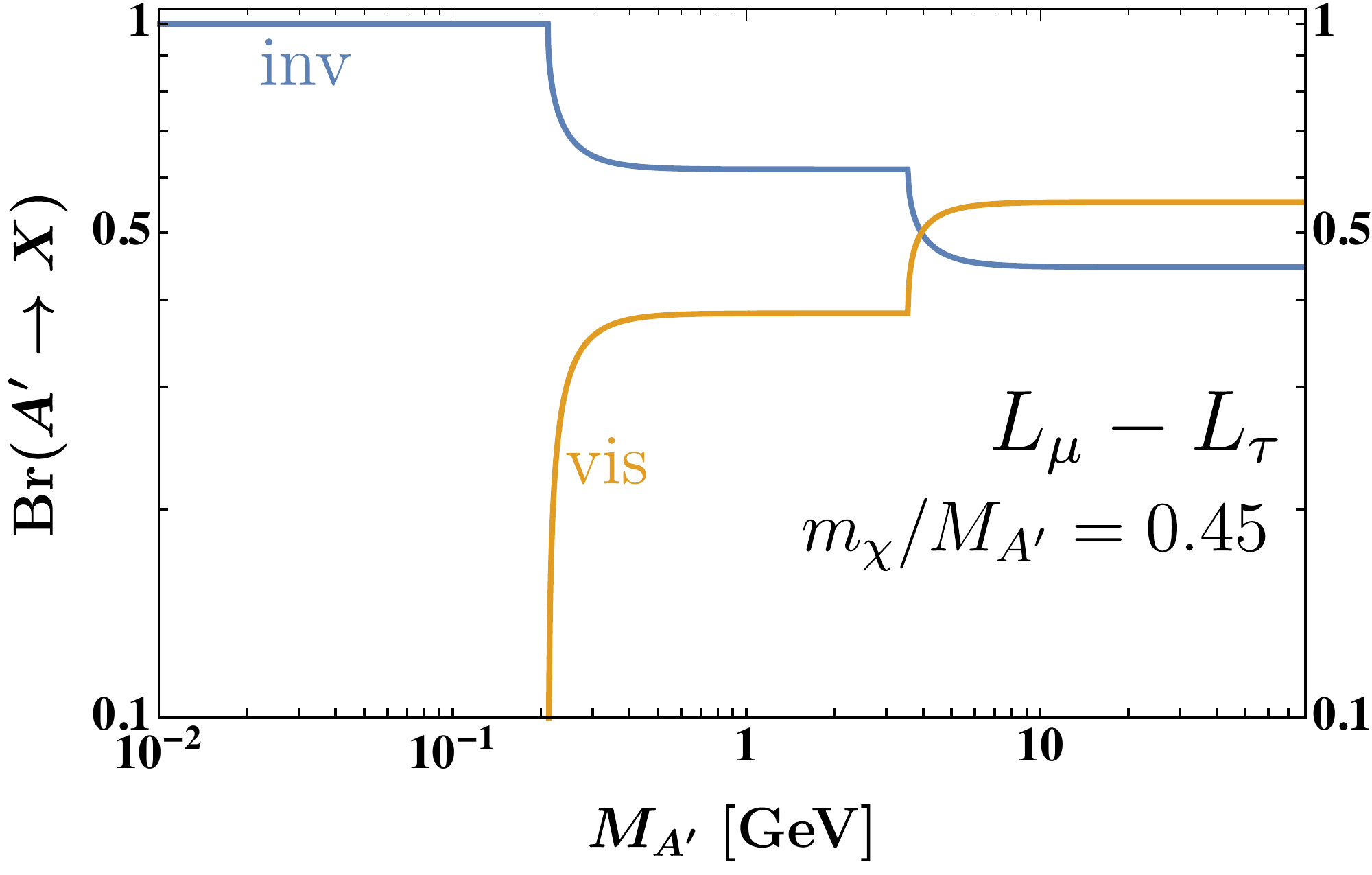}
\end{subfigure}\hfill
\begin{subfigure}{.5\textwidth}
  \centering
   \adjincludegraphics[ trim={0 0 0 0},clip, width=\textwidth]{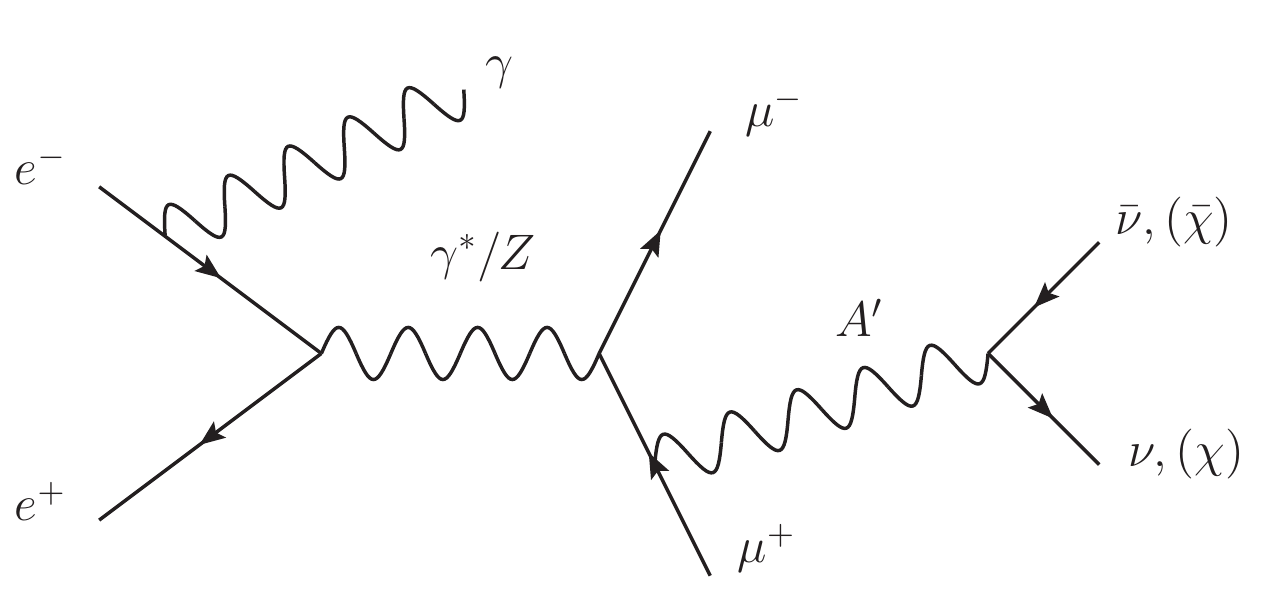}
\end{subfigure} 
 \caption{(Left) Branching ratio of the $U(1)_{L_\mu-L_\tau}$ boson $A'$ into visible (vis) versus invisible (inv) final states. (Right) Hidden photon contribution to the $\mu^+\mu^-$ + missing energy signature at an $e^+e^-$ collider. The additional tagging photon is required for reasons of background suppression.} 
\label{fig:DMsearch}
\end{figure}

From the lower panel of Figure~\ref{fig:lims_dm} it can be seen that this region could possibly be tested in searches for invisible final states in future muon runs of LDMX and NA64. These searches are especially sensitive below the dimuon threshold where the hidden photon almost always decays invisibly, either into a pair of neutrinos or DM particles $\chi$ (cf. left panel of Figure~\ref{fig:DMsearch}). In addition to these experiment, the $(g-2)_\mu$ region could potentially also be tested in a $\mu^+\mu^-$ plus missing energy signature at a low-energy $e^+e^-$ collider like e.g.~BES-III due to the process shown in the right panel of Figure~\ref{fig:DMsearch}. For experimental reasons one would require an additional tagging photon  in order to suppress the background from SM dimuon production plus missed ISR/FSR photon.

\section{Summary}

Motivated by the renormalizable vector portal interaction models of extra Abelian symmetries provide an economical but consistent framework to include new physics into the SM. The parameter space of the hidden photon of  minimal secluded $U(1)_X$ model is already quite constrained and in particular an explanation of the $(g-2)_\mu$ excess has been firmly ruled out. Beyond the minimal case, the gauge groups $U(1)_{B-L}$ and $U(1)_{L_i-L_j}$ provide for interesting phenomenology. The coupling hierarchy due to loop-induced kinetic mixing can drastically alter the sensitivity of standard hidden photon constraints to these scenarios. In particular, in the case of $U(1)_{L_\mu-L_\tau}$ an explanation of the observed $(g-2)_\mu$ is still viable. Under addition of a vector-like fermion $\chi$ the observed DM relic abundance $\Omega_\mathrm{DM}$ can be explained in the same region of parameter space. In this context,  BES-III might be able to probe the relevant region of parameter space in a search for  dimuon plus missing energy signature. Finally, searches of four-lepton final states at LHC experiments can potentially constrain the high-mass region (below the Z-pole) of the hidden photon parameter space in scenarios of gauged lepton-number differences, $U(1)_{L_i-L_j}$.

\acknowledgments

First of all, the author  wants to thank the organizers of the 7th Annual Conference on Large Hadron Collider Physics for the kind invitation and the opportunity to present this work. Furthermore, the author is very grateful to Martin Bauer and Joerg Jaeckel for collaboration on this work and to Achim Denig for stimulating discussions on invisible searches for hidden photons. The author also thanks the DFG for support via the GRK 1940 ``Particle Physics beyond the Standard Model''.

\bibliographystyle{JHEP} 
\bibliography{refs}

\providecommand{\href}[2]{#2}\begingroup\raggedright\begin{thebibliography}{10}

\bibitem{Bauer:2018onh}
M.~Bauer, P.~Foldenauer and J.~Jaeckel, \emph{{Hunting All the Hidden
  Photons}}, \href{http://dx.doi.org/10.1007/JHEP07(2018)094}{\emph{JHEP}
  \textbf{ 07} (2018) 094},
  [\href{https://arxiv.org/abs/1803.05466}{\texttt{1803.05466}}].

\bibitem{Foldenauer:2018zrz}
P.~Foldenauer, \emph{{Light dark matter in a gauged $U(1)_{L_\mu-L_\tau}$
  model}}, \href{http://dx.doi.org/10.1103/PhysRevD.99.035007}{\emph{Phys.
  Rev.} \textbf{ D99} (2019) 035007},
  [\href{https://arxiv.org/abs/1808.03647}{\texttt{1808.03647}}].

\bibitem{FoldenauerPhd}
P.~Foldenauer, \emph{{Phenomenology of Extra Abelian Gauge Symmetries}}.
\newblock PhD thesis, Heidelberg U., 2019.
\newblock
  \href{https://doi.org/10.11588/heidok.00026777}{10.11588/heidok.00026777}.

\bibitem{Hanneke:2008tm}
D.~Hanneke, S.~Fogwell and G.~Gabrielse, \emph{{New Measurement of the Electron
  Magnetic Moment and the Fine Structure Constant}},
  \href{http://dx.doi.org/10.1103/PhysRevLett.100.120801}{\emph{Phys. Rev.
  Lett.} \textbf{ 100} (2008) 120801},
  [\href{https://arxiv.org/abs/0801.1134}{\texttt{0801.1134}}].

\bibitem{Aghanim:2018eyx}
{\scshape Planck} collaboration, N.~Aghanim et~al., \emph{{Planck 2018 results.
  VI. Cosmological parameters}},
  \href{https://arxiv.org/abs/1807.06209}{\texttt{1807.06209}}.

\bibitem{Okun:1982xi}
L.~B. Okun, \emph{{LIMITS OF ELECTRODYNAMICS: PARAPHOTONS?}}, {\emph{Sov. Phys.
  JETP} \textbf{ 56} (1982) 502}.

\bibitem{Holdom:1985ag}
B.~Holdom, \emph{{Two U(1)'s and Epsilon Charge Shifts}},
  \href{http://dx.doi.org/10.1016/0370-2693(86)91377-8}{\emph{Phys. Lett.}
  \textbf{ 166B} (1986) 196--198}.

\bibitem{ArkaniHamed:2008qn}
N.~Arkani-Hamed, D.~P. Finkbeiner, T.~R. Slatyer and N.~Weiner, \emph{{A Theory
  of Dark Matter}},
  \href{http://dx.doi.org/10.1103/PhysRevD.79.015014}{\emph{Phys. Rev.}
  \textbf{ D79} (2009) 015014},
  [\href{https://arxiv.org/abs/0810.0713}{\texttt{0810.0713}}].

\bibitem{ArkaniHamed:2008qp}
N.~Arkani-Hamed and N.~Weiner, \emph{{LHC Signals for a SuperUnified Theory of
  Dark Matter}},
  \href{http://dx.doi.org/10.1088/1126-6708/2008/12/104}{\emph{JHEP} \textbf{
  12} (2008) 104},
  [\href{https://arxiv.org/abs/0810.0714}{\texttt{0810.0714}}].

\bibitem{Hooper:2008im}
D.~Hooper and K.~M. Zurek, \emph{{A Natural Supersymmetric Model with MeV Dark
  Matter}}, \href{http://dx.doi.org/10.1103/PhysRevD.77.087302}{\emph{Phys.
  Rev.} \textbf{ D77} (2008) 087302},
  [\href{https://arxiv.org/abs/0801.3686}{\texttt{0801.3686}}].

\bibitem{Baumgart:2009tn}
M.~Baumgart, C.~Cheung, J.~T. Ruderman, L.-T. Wang and I.~Yavin,
  \emph{{Non-Abelian Dark Sectors and Their Collider Signatures}},
  \href{http://dx.doi.org/10.1088/1126-6708/2009/04/014}{\emph{JHEP} \textbf{
  04} (2009) 014},
  [\href{https://arxiv.org/abs/0901.0283}{\texttt{0901.0283}}].

\bibitem{Cheung:2009qd}
C.~Cheung, J.~T. Ruderman, L.-T. Wang and I.~Yavin, \emph{{Kinetic Mixing as
  the Origin of Light Dark Scales}},
  \href{http://dx.doi.org/10.1103/PhysRevD.80.035008}{\emph{Phys. Rev.}
  \textbf{ D80} (2009) 035008},
  [\href{https://arxiv.org/abs/0902.3246}{\texttt{0902.3246}}].

\bibitem{Katz:2009qq}
A.~Katz and R.~Sundrum, \emph{{Breaking the Dark Force}},
  \href{http://dx.doi.org/10.1088/1126-6708/2009/06/003}{\emph{JHEP} \textbf{
  06} (2009) 003},
  [\href{https://arxiv.org/abs/0902.3271}{\texttt{0902.3271}}].

\bibitem{Morrissey:2009ur}
D.~E. Morrissey, D.~Poland and K.~M. Zurek, \emph{{Abelian Hidden Sectors at a
  GeV}}, \href{http://dx.doi.org/10.1088/1126-6708/2009/07/050}{\emph{JHEP}
  \textbf{ 07} (2009) 050},
  [\href{https://arxiv.org/abs/0904.2567}{\texttt{0904.2567}}].

\bibitem{Batell:2009yf}
B.~Batell, M.~Pospelov and A.~Ritz, \emph{{Probing a Secluded U(1) at
  B-factories}},
  \href{http://dx.doi.org/10.1103/PhysRevD.79.115008}{\emph{Phys. Rev.}
  \textbf{ D79} (2009) 115008},
  [\href{https://arxiv.org/abs/0903.0363}{\texttt{0903.0363}}].

\bibitem{Essig:2009nc}
R.~Essig, P.~Schuster and N.~Toro, \emph{{Probing Dark Forces and Light Hidden
  Sectors at Low-Energy e+e- Colliders}},
  \href{http://dx.doi.org/10.1103/PhysRevD.80.015003}{\emph{Phys. Rev.}
  \textbf{ D80} (2009) 015003},
  [\href{https://arxiv.org/abs/0903.3941}{\texttt{0903.3941}}].

\bibitem{Reece:2009un}
M.~Reece and L.-T. Wang, \emph{{Searching for the light dark gauge boson in
  GeV-scale experiments}},
  \href{http://dx.doi.org/10.1088/1126-6708/2009/07/051}{\emph{JHEP} \textbf{
  07} (2009) 051},
  [\href{https://arxiv.org/abs/0904.1743}{\texttt{0904.1743}}].

\bibitem{Bjorken:2009mm}
J.~D. Bjorken, R.~Essig, P.~Schuster and N.~Toro, \emph{{New Fixed-Target
  Experiments to Search for Dark Gauge Forces}},
  \href{http://dx.doi.org/10.1103/PhysRevD.80.075018}{\emph{Phys. Rev.}
  \textbf{ D80} (2009) 075018},
  [\href{https://arxiv.org/abs/0906.0580}{\texttt{0906.0580}}].

\bibitem{Bennett:2006fi}
{\scshape Muon g-2} collaboration, G.~W. Bennett et~al., \emph{{Final Report of
  the Muon E821 Anomalous Magnetic Moment Measurement at BNL}},
  \href{http://dx.doi.org/10.1103/PhysRevD.73.072003}{\emph{Phys. Rev.}
  \textbf{ D73} (2006) 072003},
  [\href{https://arxiv.org/abs/hep-ex/0602035}{\texttt{hep-ex/0602035}}].

\bibitem{Sirunyan:2018nnz}
{\scshape CMS} collaboration, A.~M. Sirunyan et~al., \emph{{Search for an
  $L_{\mu}-L_{\tau}$ gauge boson using Z$\to4\mu$ events in proton-proton
  collisions at $\sqrt{s} =$ 13 TeV}},
  \href{http://dx.doi.org/10.1016/j.physletb.2019.01.072}{\emph{Phys. Lett.}
  \textbf{ B792} (2019) 345--368},
  [\href{https://arxiv.org/abs/1808.03684}{\texttt{1808.03684}}].

\bibitem{Parker:2018vye}
R.~H. Parker, C.~Yu, W.~Zhong, B.~Estey and H.~M{\"ul}ler, \emph{{Measurement
  of the fine-structure constant as a test of the Standard Model}},
  \href{http://dx.doi.org/10.1126/science.aap7706}{\emph{Science} \textbf{ 360}
  (2018) 191}, [\href{https://arxiv.org/abs/1812.04130}{\texttt{1812.04130}}].

\bibitem{Heeck:2018nzc}
J.~Heeck, M.~Lindner, W.~Rodejohann and S.~Vogl, \emph{{Non-Standard Neutrino
  Interactions and Neutral Gauge Bosons}},
  \href{http://dx.doi.org/10.21468/SciPostPhys.6.3.038}{\emph{SciPost Phys.}
  \textbf{ 6} (2019) 038},
  [\href{https://arxiv.org/abs/1812.04067}{\texttt{1812.04067}}].

\end{thebibliography}\endgroup

\end{document}